\documentclass[11pt,a4paper]{article}

\usepackage{color,colortbl}
\definecolor{hellblau}{rgb}{0,0.7,1}
\definecolor{weis}{rgb}{0,0,0}
\usepackage[latin1]{inputenc}
\usepackage[bindingoffset=0.5cm,right=2.25cm]{geometry}
\usepackage{graphicx}
\usepackage[official]{eurosym}
\usepackage{booktabs}
\usepackage{caption}
\usepackage{setspace}
\usepackage{float}
\usepackage{url}
\usepackage{rotating}  
\usepackage{braket}
\usepackage{makeidx}
\usepackage{rccol}
\usepackage{fltpoint}
\usepackage{dcolumn}
\newcolumntype{k}{D{,}{,}{5,2}}
\usepackage{verbatim} 
\usepackage{natbib}
\usepackage{amsmath}
\usepackage[bookmarksnumbered=true,bookmarksopen=true]{hyperref}

\usepackage{url}
\newcommand{\expectation}[3][0]{%
  \ifcase#1
     E( #2 \mid #3 )
     \or E \bigl( #2 \bigm\vert #3 \bigr)
     \or E \Bigl( #2 \Bigm\vert #3 \Bigr)
     \or E \biggl( #2 \biggm\vert #3 \biggr)
     \or E \Biggl( #2 \Biggm\vert #3 \Biggr)
  \else
     E \left( #2  \;\middle\vert\; #3 \right)
  \fi}

\usepackage{lscape} 

\author{Stefan Kerbl\footnote{Oesterreichische Nationalbank (OeNB), email: stefan.kerbl@oenb.at. \newline This article should not be reported as representing the views of the OeNB. The views expressed here are those of the author. I am grateful to Stefan Theussl and Maria Oberleithner for support in acquiring computing power.}} 

\title{\vspace{-0.8cm} \Large{Regulatory Medicine Against Financial Market Instability:\\What Helps And What Hurts?}\\[0.4cm]\Large{}}   
\begin{document}

\date{Version 1.5 \\ 2010}

\maketitle
\setlength{\baselineskip}{1.1\baselineskip}

\begin{center}
\textbf{\large{Abstract}}
\end{center}  %
Do we know if a short selling ban or a Tobin Tax result in more stable asset prices? Or do they in fact make things worse? Just like medicine regulatory measures in financial markets aim at improving an already complex system. And just like medicine these interventions can cause side effects which are even harder to assess when taking the interplay with other measures into account.
In this paper an agent based stock market model is built that tries to find answers to the questions above. 
In a stepwise procedure regulatory measures are introduced and their implications on market 
liquidity and stability examined. Particularly, the effects of \textit{(i)} a ban of short selling \textit{(ii)} a mandatory risk limit, i.e. a Value-at-Risk~limit, \textit{(iii)} an introduction of a Tobin~Tax, i.e. transaction tax on trading, and \textit{(iv)} any arbitrary combination of the measures are observed and discussed. 
The model is set up to incorporate non-linear feedback effects of  leverage and liquidity constraints leading to fire sales and escape dynamics. 
In its unregulated version the model outcome is capable of reproducing stylised facts of asset returns like fat tails and clustered volatility. 
Introducing regulatory measures  shows that only a mandatory risk limit is beneficial from every perspective, while a short selling ban --- though reducing volatility --- increases  tail risk. The contrary holds true for a Tobin Tax: it reduces the occurrence of crashes but increases volatility. Furthermore, the interplay of measures is not negligible: measures block each other and a well chosen combination can mitigate unforeseen side effects. Concerning the Tobin Tax the findings indicate that an overdose can do severe harm.
\\

\\

\setlength{\baselineskip}{1.1\baselineskip}

\section{Motivation} \label{motivation}

\begin{quotation}
``It's remarkable that while any
new technical device or medical drug has
extensive testing for efficiency, reliability and
safety before it ever hits the market, we still
implement new economic measures without
any prior testing.'' 

\hspace{-0.5 cm} --- Dirk Helbing, Swiss Federal Institute of Technology Zurich in \cite{buchanan-2009}.
\end{quotation}


The financial crisis spurred the discussion about further regulations in asset markets. However, the consequences of imposing a transaction tax, a short selling ban or mandatory risk limits are unknown  to a large extent. ``Prior testing'' is both: hardly feasible and absolutely necessary. Hardly feasible, because the large sums of money handled, the interconnectedness of actions etc. do not allow for lab experiments. Only within newly emerged  agent based models this task seems doable. 
At the same time, prior testing is absolutely necessary. Like the human body the financial market is an enormously complex organism and like medicine regulatory measures aim at improving it. Side effects or unforeseen interactions of measures require prior testing, as the quotation above demands. This paper provides  evidence if indeed imposing regulatory measures makes markets more stable. To do so, an agent based model framework is set up, which shares basic ideas of \cite{thurner-2009}. Subsequently, this baseline model is modified by the introduction of regulatory measures. \\

\textbf{\textit{Why Agent Based Models?}}

The dynamics of financial markets pose a challenge to research just as they pose a threat to financial stability. Typically, asset returns are characterised by so called stylised facts including fat tails and clustered volatility (see \citealt{rama-2001}). Classic economic theory fails to predict such behaviour. Recent literature, however, has shown that by the incorporation of leverage, fire sales, escape dynamics and liquidity constraints stylised facts occur (\citealt{friedman-2009}). 
The financial crisis confirmed the importance of taking such effects into account. However,  modelling of  interaction effects in an analytical framework soon becomes untractable.
Therefore, the dramatic increase of computational power over the past decades gave rise to agent based models. Within such models one is allowed to move away from  the classical modelling approach featuring the representative agent but to model the action of each and every actor, thus integrating non-linear feedback dynamics.\footnote{Numerous papers give witness to the popularity agent based models gained over the past two decades, see \cite{LeBaron-2001} and \cite{LeBaron-2006} for extended literature discussions.}\\

While typical agent based models feature heterogeneous agents who dynamically optimize seemingly irrational strategies\footnote{Compare for instance trend followers in models of \cite{Lux-1998} and \cite{deGrauwe-2006a} or the Minority Game literature, e.g. \cite{challet-2001} and \cite{Sornette-2007}.},  \cite{thurner-2009} recently showed that even under the assumption of relatively rational value-investors fat tails occur when feedback effects of leverage are incorporated. 
The baseline model introduced in Section \ref{baseline} draws on \cite{thurner-2009}. It models leveraged agents who trade a single asset according to a mispricing signal. In its unregulated version the model reproduces fat tails and clustered volatility.\\

\textbf{\textit{Regulatory Measures}}

The focus will then be shifted to the question of interest, the impact of regulatory measures. The financial crisis has amplified voices demanding a stronger regulatory framework of asset markets. Among the cloud of demands the following are picked for closer examination: 

\begin{itemize}
	\item[\textit{(i)}] a ban of short selling, 
	\item[\textit{(ii)}] a mandatory risk limit and
	\item[\textit{(iii)}] a Tobin Tax, i.e. transaction tax on trading\footnote{While the Tobin Tax was originally  suggested only for foreign exchange rate markets, the term is now regularly applied to mean a tax on financial transactions in general. This paper will use the terms transaction tax and Tobin Tax synonymously.}.
\end{itemize}

In each of the three cases a high level of uncertainty concerning the consequences of an introduction prevent  a fact-led discussion. This is probably best seen by reading the following two citations of U.S. Securities and Exchange Commission Chairman Christopher Cox. The first quote was said  at the time of introduction of the short selling ban in US-stock markets in September 2008 (\citealt{NYT-2008}) and the second only three months later in December 2008 (\citealt{Reuters-2008}):

\begin{quote}
``The emergency order temporarily banning short selling of financial stocks will restore equilibrium to markets.''
\end{quote}

\begin{quote}
``While the actual effects of this temporary action will not be fully understood for many more months, if not years, knowing what we know now, I believe on balance the commission would not do it again. \dots{} The costs appear to outweigh the benefits.''
\end{quote}


In their empirical study \cite{Marsh-2008} find ``no strong evidence that (short selling bans) have been effective in reducing share price volatility or limiting share price falls.'' Further  studies based on observed data (e.g. \citealt{Lobanova-2009}, \citealt{Boehmer-2009} and \citealt{Beber-2009}) find rather negative effects of  short selling restrictions on market liquidity and increasing effects on volatility.\\


Similarly, the adoption of a Tobin Tax, i.e. transaction tax on trading, has lead to controversy within the field of  academics as well as within politics. Originally proposed by \cite{Tobin-1978}, the tax now enjoys great popularity as a potential means to reduce market volatility and as source for tax revenues. In fact, merely naming supporters and opposers of the tax would be way out of the scope of this paper. However, a clear reflection of the popularity can be grasped by the length of the respective article in \cite{wiki}, which also provides a comprehensive list of the numerous supporters and opposers in politics. In the academic world, studies come to mixed conclusions. While a negative effect on trading volume is generally agreed upon, the impact on price volatility 
is less clear cut and even contrary, leading   \cite{Hanke-2007} to infer that ``in sum, the literature on the effects of a Tobin tax on market efficiency arrives at opposite ends. \dots{} there is no general agreement on the consequences of a Tobin tax on price volatility.'' While some argue that a transaction tax reduces the trading of rather uninformed actors, thus leading to more efficient and less volatile markets, others argue that a transaction tax prevents flexible price adjustment to new information and therefore rather leads to price jumps and higher volatility (see also the debate in \citealt{Hanke-2007}). \cite{Westerhoff-2003}, \cite{Westerhoff-2005}, \cite{Westerhoff-2007} and \cite{Mannaro-2008} study the imposition of a transaction tax within the framework of agent based models. While the latter conclude that volatility rises with the imposition of a Tobin Tax, the other papers find that the effects depend on the liquidity of a market and on the magnitude of the tax.\\

The third regulatory measure, titled mandatory risk limits,  may seem less debated, but is in fact  already in place for many of the larger market participants like banks via the Basel regime. 
In such a regime, agents are obliged to quantify their risk and relate this risk  to 
 their own funds, thus keeping their theoretical default probability below a certain threshold. 
 Insurers and hedge funds are as well required to run risk managements techniques -- a regulation that is currently intensified.  
 Irrespective of the regulatory framework, risk quantification and risk limiting has become a general practice among the major market participants, i.e. funds and banks (see e.g. chapter 1.1 in \citealt{McNeil-2005}). While a sound risk management is without doubt for the benefit of the single institution, its consequences for systemic risks are ambiguous. To see this, imagine an agent close to its risk limit when stocks decline. The decline not only shrinks her own funds but may also increases the risk quantified for the same position. This may in turn lead to fire sales, thus amplifying the initial shock. Such phenomena combined with strategy herding could potentially lead to severe downturn momentum.\footnote{Strategy herding is in fact a major driver for market crashes in agent based Minority Games. See e.g. \cite{Sornette-2007}.}\\

This paper not only discusses the implications of the three regulatory measures, but further provides evidence on their potential interplay. 
 While the interplay of drugs and their side effects is a pervasive topic in medical research, it is much less debated in the context of financial markets. Thus, this paper aims at giving answers to the interplay of the regulatory measures. \\

To conclude, the contribution of the paper is threefold. Firstly, while the effects of a short selling ban and of a transaction tax have already been studied, there remains a level of uncertainty that requires further research. Furthermore, the effects of risk limits --- though beneficial on the individual level --- may have negative side effects, which seem to have been neglected in the scientific discussion. Secondly, existing literature approaches the questions usually either from an empirical view using observed data or a reduced form theoretical model, but not within the framework of an agent based model\footnote{Note the exceptions concerning the Tobin~Tax: \cite{Westerhoff-2003}, \cite{Westerhoff-2005}, \cite{Westerhoff-2007} and \cite{Mannaro-2008}.}. Thirdly, to the author's knowledge this paper is first in examining the combination of these regulatory measures. \\

The remainder of the paper is structured as follows. Section \ref{baseline} introduces the baseline model with no regulations in place. Subsequently, Section \ref{regulation} presents adjustments due to the regulatory measures. Finally, Section \ref{results} discusses the results while Section \ref{conclusions} concludes and outlines potential shortcomings of the approach, thus suggesting ways of further research.

\section{The Baseline Model} \label{baseline}

This section describes the baseline model representing the unrestricted market. The description  starts at the most general level and successively works downwards explaining the model in more detail.\\

At the top level there is the market clearing equation defining that at each timestep $t$ total demand, as sum over the $N^{a}$ individual demands $D^{}_{i, t}$,  must equal the total number of shares $N^{s}$, therefore ensuring that supply meets demand and the market clears.

\begin{equation} \label{eq-price}
	\sum^{N^{a}}_{i=1} D^{}_{i, t}(p_{t})= N^{s}
\end{equation}

As described below, demand of each single agent is a function of price $p_{t}$ among others. By solving Equation~(\ref{eq-price})  one obtains the price.\footnote{This set--up is  more sophisticated than the one used by usual agent based models, in which price is a (linear) function of ``excess demand'', which implies a linear response to market movements (e.g. \citealt{friedman-2009}). It comes, however, at the cost of more complex computational demands.} At each timestep agents choose  the fraction of their total wealth $W^{}_{i,t}$ to be invested in cash $C^{}_{i,t}$ and in  shares, therefore 

\begin{equation} \label{eq-wealth}
	W^{}_{i,t} = C^{}_{i,t}+ p_{t} \, D^{}_{i, t}(p_{t}).
\end{equation}

Before turning to the demand equations, note that when $D^{}_{i,t}<0$ agents take a short position and when $D^{}_{i,t}>0$ they are long. To fund their actions agents can leverage themselves up to a maximum leverage of $\lambda^{max}$.\footnote{According to modern standard, leverage is defined as the asset side (of the balance sheet) divided by own funds, therefore $\lambda^{long}_{i,t}:=p_{t} \, D^{}_{i, t}(p_{t})/W^{}_{i,t}$ and $\lambda^{short}_{i,t}:= (W^{}_{i,t}- p_{t}\,D^{}_{i, t}(p_{t}))/W^{}_{i,t}$.}  
As long as leverage is not at its maximum, agents' demand is a linear function of the perceived mispricing signal. This mispricing signal is the difference between the current price and the perceived fundamental value, thus $m_{i,t}:=p_{t}-p^{perc}_{i,t}$.  This leads to the demand functions:

\begin{equation}  
\label{eq-demand}
	D^{}_{i,t} = \begin{cases}
 	 (1-\lambda^{max}) \, W^{}_{i,t} / p_{t}    & \quad \text{if} \quad m_{i,t}<m^{crit,short}_{i,t} \\
	 \lambda^{max} \, W^{}_{i,t} / p_{t}        & \quad \text{if} \quad m_{i,t}>m^{crit,long}_{i,t} \\
	 \beta_{i}\, m_{i,t} \, W^{}_{i,t} / p_{t}  & \quad \text{otherwise,}
\end{cases}
\end{equation}  

where $\beta_{i}$ represents a parameter denoting the aggressiveness of the agent, that is how fast he reacts to price signals and $m^{crit}_{i,t}$  the mispricing signal which would lead to the use of the maximum leverage.   Thus, $m^{crit,short}_{i,t}=(1-\lambda^{max})/\beta_{i}$ if agent $i$ is in a short position and $m^{crit,long}_{i,t}=\lambda^{max}/\beta_{i}$ if she is long. While the first two lines of Equation~(\ref{eq-demand}) simply limit the demand to its maximum leverage, the third specifies demand in the unbounded case as a linear function of the mispricing signal and the aggressiveness of the agent, $\beta_{i}$.\footnote{In fact, the underlying utility function would be (subscripts omitted): $U(D,C)=D^{\beta \, m} \, C^{1-\beta}$. See also  \cite{thurner-2009}.}
Price and wealth in Equation~(\ref{eq-demand}) ensures that at a given mispricing signal two equally aggressive agents will invest the same fraction of their wealth.\footnote{\label{f-tau}As in practice only a fraction of agents actually take short positions, for simulation purpose define $\tau$ as the fraction of agents who avoid taking short positions even in the baseline model.}\\

Until now, the perceived fundamental value of the share, $p^{perc}_{i,t}$, was left unspecified. In this model, agents' perceptions follows a discrete Ornstein-Uhlenbeck process that guarantees the perceived values to be wander around but mean revert to the fundamental value. 

\begin{equation} \label{eq-perc}
\log p^{perc}_{i,t}=  \rho \, \log p^{perc}_{i,t-1} +  (1-\rho ) \, \log V  + \epsilon_{i,t},        
\end{equation} 
where $V=1$ denotes the \textit{true} fundamental value, $\epsilon \sim N(0,\Sigma)$
and $0<\rho<1$. In order to mirror 
 market wide misjudgement and herding $\epsilon$ correlates across agents. 
Finally, in each round $t$ before each market participant $i$ computes his demand according to Equation~(\ref{eq-demand}) and the price $p_{t}$ is derived according Equation~(\ref{eq-price})\footnote{Note that Equation~(\ref{eq-demand}) and Equation~(\ref{eq-price}) have to be solved simultaneously as both depend on the other.} the wealth $W_{i,t}$ is updated according to

\begin{equation} \label{eq-update}
	W^{}_{i,t} = W^{}_{i,t-1}+  D^{}_{i, t-1}\,(p_{t}-p_{t-1} ). 
\end{equation}

 
In line with \cite{thurner-2009}, agents default if their wealth, $W_{i,t}$, decreases below 10\% of their initial wealth and are reintroduced after 100 timesteps. \\

\begin{figure}
			\begin{center}
			\includegraphics[width=0.85\textwidth]{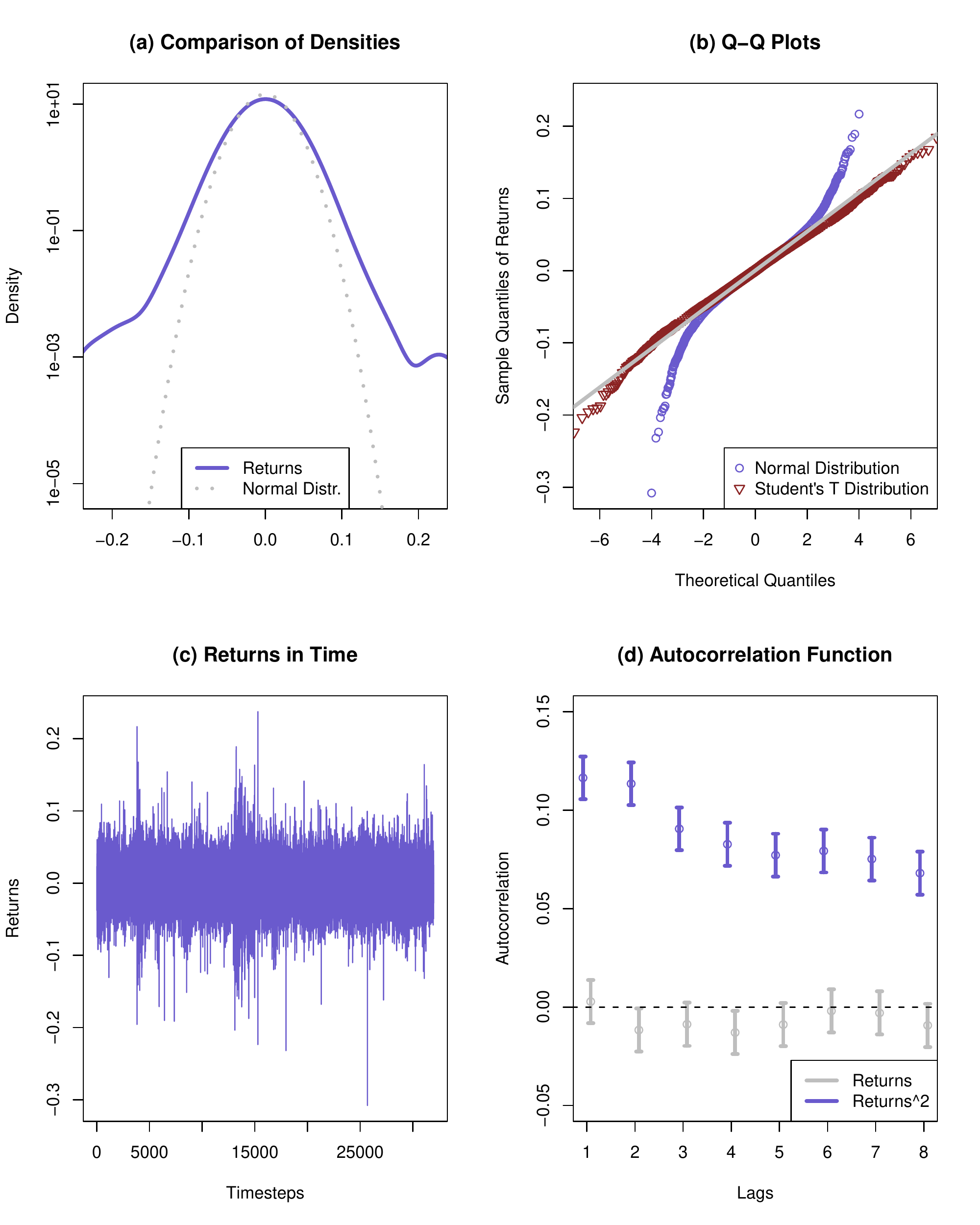}
			\end{center}
				\caption{Statistical Properties of Returns: (a)~display of kernel density estimates of the returns compared to the normal distribution with respective standard deviation, (b)~Q-Q plot of returns against normal and student t~distributions with 7~df, (c)~plot of a typical draw of returns, (d)~autocorrelation function for returns and squared returns with 95\% confidence intervalls.}
				\label{supi}
			\end{figure}

Figure~\ref{supi} displays the implied characteristics of the returns of the unregulated model, calibrated according to 
Table~\ref{tab_calibration} (see appendix, page~\pageref{tab_calibration}). While plot \textit{a} and \textit{b} display  excess kurtosis present in the implied time series of returns as well as a gain/loss asymmetry, plot \textit{c} and \textit{d} provide evidence on the absence of autocorrelation among returns but non-zero autocorrelation among squared returns, i.e. clustered volatility is present (see \citealt{rama-2001}). The emergence of these characteristics are endogenous considering the normal iid distribution of $\vec{\epsilon_{t}}$ in Equation~(\ref{eq-perc}).

\section{Short selling ban, risk limits and transaction tax} \label{regulation}

Having specified the unregulated model, this section presents the amendments for each regulatory measure in sequence.\\

\textbf{\textit{Short selling ban}}

In the unregulated market, agents' demand of shares can be positive or negative alike. In the latter case, the agent goes short. To implement a short selling ban the demand Equation~(\ref{eq-demand}) has to be adjusted to cap the demand at zero. 

\begin{equation}  
\label{eq-demand-short}
	D^{}_{i,t} = \begin{cases}
 	 0    																			& \quad \text{if} \quad m_{i,t}\leq 0 \\
	 \lambda^{max} \, W^{}_{i,t} / p_{t}        & \quad \text{if} \quad m_{i,t}>m^{crit,long}_{i,t} \\
	 \beta_{i}\, m_{i,t} \, W^{}_{i,t} / p_{t}  & \quad \text{otherwise.}
\end{cases}
\end{equation}  

Note that in comparison to Equation~(\ref{eq-demand}) only the first line changed. \\

\textbf{\textit{Value-at-Risk limits}}

As discussed in the introduction, Value-at-Risk is now a widely applied concept in risk management. Hereby, one quantifies the risk of a given position according to a quantile of the  estimated loss distribution. While different methods are applied in practice, this paper sticks to the popular and straight forward variant called variance-covariance approach. Hence, at each timestep market participants calculate their individual Value-at-Risk for holding one unit of the asset: 

\begin{equation} \label{eq-var}
	\text{VaR}_{i,t}= \mu_{i,t}-\alpha*\sigma_{i,t},
\end{equation}

where $\mu_{i,t}$ and $\sigma_{i,t}$ are empirical estimates of mean and standard deviation of asset returns which might deviate among agents and $\alpha=\Phi^{-1}(0.99)$ represents the 99\%-quantile of the normal distribution. Agents subjected to a Value-at-Risk~limit are  not only bound by the maximum leverage constraint but by a maximum portfolio Value-at-Risk as well, which aims at reducing the default probability of agents below a certain threshold. To adjust the demand equation of the baseline model, define $m^{crit, var}_{i,t}:=(\beta_{i}\,\text{VaR}_{i,t})^{-1} $ as the critical mispricing signal, at which the unbounded demand would be higher than the maximum Value-at-Risk.\footnote{The deviation is simple, considering that the Value-at-Risk concept limits the portfolio VaR, $D^{}_{i,t}\,\cdot\, p_{t}\,\cdot\,\text{VaR}_{i,t}$, to equal $W^{}_{i,t}$ at maximum.} Consequently, Equation~(\ref{eq-demand}) changes to

\begin{equation}  
\label{eq-demand-var}
	D^{}_{i,t} = \begin{cases}
 	 (1-\lambda^{max}) \, W^{}_{i,t} / p_{t}    & \quad \text{if} \quad m_{i,t}<m^{crit,short}_{i,t} \\
	 - W^{}_{i,t} / (p_{t}\,\text{VaR}_{i,t}) 						& \quad \text{if} \quad m_{i,t}<-m^{crit,var}_{i,t} \\
	 \lambda^{max} \, W^{}_{i,t} / p_{t}         & \quad \text{if} \quad m_{i,t}>m^{crit,long}_{i,t} \\
	 W^{}_{i,t} / (p_{t}\,\text{VaR}_{i,t})						& \quad \text{if} \quad m_{i,t}>m^{crit,var}_{i,t}   \\ 
	 \beta_{i}\, m_{i,t} \, W^{}_{i,t} / p_{t}  & \quad \text{otherwise,}
\end{cases}
\end{equation}  

while in case more than one  restriction hits, the one that satisfies $\min(|D^{}_{i,t}|)$ is in effect. Comparing the baseline model of Equation~(\ref{eq-demand}) with Equation~(\ref{eq-demand-var})  above one finds that simply two new lines have emerged holding the implied risk in check. Analogously, the simultaneous reign of a short selling ban and a Value-at-Risk~limit, would bind the demand to zero if $m_{i,t}\leq0$, while the remainder of Equation~(\ref{eq-demand-var}) would hold.\\

\textbf{\textit{Transaction Tax}}

In its core, a transaction tax reduces the expected return of an investment by (twice)\footnote{The tax is applied at buying and selling the asset.} the  tax level applied. Agents will therefore require a higher expected payoff for the same level of investment, i.e. their demand in the asset. 
 Consequently, agents will  keep their current demand unchanged, if they are subjected to only a minor mispricing signal, or in other words will only change their demand if the mispricing signal is strong enough, so that expected payoffs of the trade are positive. Hence, under the regime of a transaction tax, agents compute their demand 
 according to 

\begin{equation}  
\label{eq-demand-tobin}
	D^{}_{i,t} = \begin{cases}
 	 D^{}_{i,t-1}																& \quad \text{if} \quad |D^{*}_{i,t}-D^{}_{i,t-1}|<\Gamma \\
 	 (1-\lambda^{max}) \, W^{}_{i,t} / p_{t}    & \quad \text{if} \quad m_{i,t}<m^{crit,short}_{i,t} \\
	 \lambda^{max} \, W^{}_{i,t} / p_{t}        & \quad \text{if} \quad m_{i,t}>m^{crit,long}_{i,t} \\
	 \beta_{i}\, m_{i,t} \, W^{}_{i,t} / p_{t}  & \quad \text{otherwise,}
\end{cases}
\end{equation}

where  $\Gamma$ is a threshold for the mispricing signal, and $D^{*}_{i,t}$ is the demand that would result without incorporation of the Tobin Tax, i.e. without the first line of Equation~(\ref{eq-demand-tobin}). Note that only this first line is new and  in case more than one  restriction hits, the one that satisfies $\min(|D^{}_{i,t}|)$ is in effect. This ensures that changes in demand due to shifts in the  mispricing signal,  $m_{i,t}$, require a defined magnitude. In the simulation the threshold is chosen in order to keep (empirically determined) expected returns of trading positive.\\

\begin{samepage} 
\textbf{\textit{Combination of Regulatory Measures}} \nopagebreak

For any arbitrary combination of the three regulatory measures one has to merge the formulas from above. As in each case restrictions are added to the original demand, merging them is straight forward. For instance, under a short selling ban and a Value-at-Risk~limit demand is limited to zero or positive values, while at the upper bound the leverage and Value-at-Risk~limit bind demand against becoming excessive. 
 As the final set of equation for each combination of regulatory measures is somewhat lengthy their display is omitted here.
\end{samepage}

\include{calibration}

\section{Results} \label{results}

This section will now turn to the simulated impacts of the regulatory measures on \textit{(i)} 
 market liquidity, \textit{(ii)} market volatility, \textit{(iii)} market stability, i.e. the risk of tail events, and \textit{(iv)} probability of default of an agent. 
%
%
A reasonable estimate for market liquidity seems to be
\begin{equation}
	\text{\textit{liquidity}}:=\frac{1}{(N^{t}-1)\, N^{a}}\sum^{N^{t}}_{t=2} \sum^{N^{a}}_{i=1}  |D_{i,t}-D_{i,t-1}| \, ,
\end{equation}
i.e. the average amount of shares traded by an agent per timestep. $N^{t}$ denotes the number of timesteps in one simulation run.
%
 To measure market volatility the standard deviation of returns will be evaluated and for market stability its respective (excess) kurtosis, as a measure for extreme shifts in the price.\footnote{As large upswings do not pose a thread to financial stability in themselves, the kurtosis is evaluated from the distribution in which negative returns are flipped at zero and positive ones are ignored. It turned out that none of the results depend on the flipping and the new measure is correlated by more than 0.89 with the standard kurtosis.} Finally, the number of defaults of each run is evaluated.\\

\begin{figure}
			\begin{center}
			\includegraphics[height=0.9\textheight,keepaspectratio]{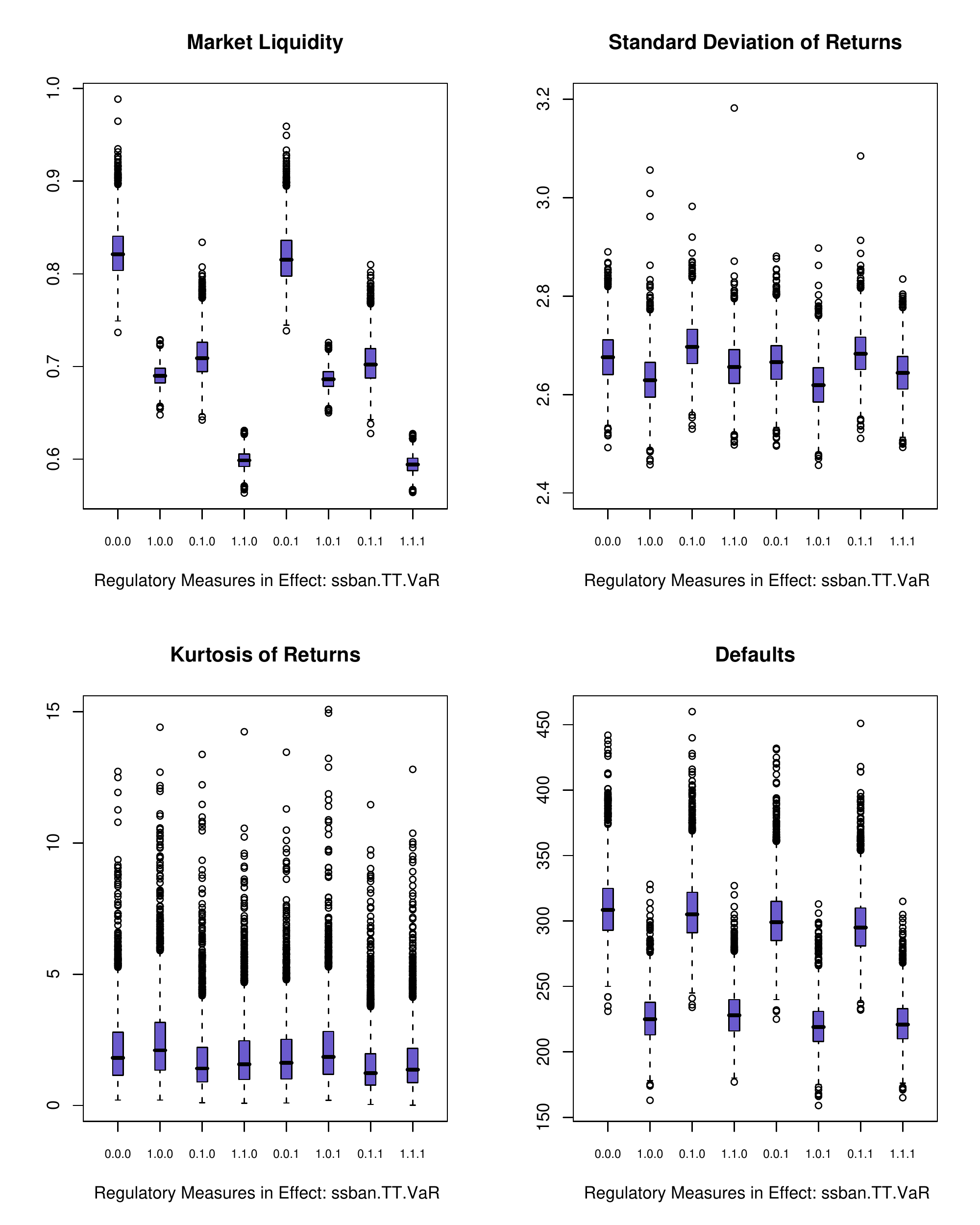}
			\end{center}
				\caption{Distributions of market characteristics under different regulatory regimes.}
				\label{box1}
			\end{figure}

As a first result, figure~\ref{box1} 
visualises the resulting  distribution of the relevant metrics. 
 Market liquidity shows strong dependence on  both short selling restrictions and the Tobin~Tax. Not only does market liquidity drop, also its volatility across runs is drastically reduced. 
 The volatility of returns seems to be negatively affected by a short selling ban, while also a mandatory VaR~limit contributes. Concerning the kurtosis, there is less clear cut evidence. Its distributions is  strongly skewed to the right -- in fact, a few  points lie far in the extreme tail, even overreaching 50. The number of defaults is obviously affected by a short selling ban, which reduces their numbers strongly. \\

To  assess the effects of regulatory measures more closely liquidity, 
standard deviation, kurtosis and number of defaults 
are regressed on the exogenous dummy variables short selling ban, Value-at-Risk~limit, Tobin~Tax and respective interaction terms as  indicating the regulatory measure to be in effect or not. As indicated by figure~\ref{box1} -- especially the liquidity plot -- , heteroskedasticity is an issue. Hence, the regression was conducted using feasible GLS\footnote{Alternatively, one can look at the quantiles of the resulting distributions. Find the respective  quantile regression results in the appendix, Table~\ref{tab_quantreg}, page~\pageref{tab_quantreg}. In short, the  picture modelling the median is  very similar both in terms of magnitude and significance of the parameters compared to the feasible GLS regression results displayed in this section.}. Table~\ref{tab_results} displays the regression results with stars indicating statistical significance. With an adjusted R squared of 0.89 the liquidity model manages to explain a relatively high fraction of the endogenous variance. Interestingly, all regulatory measures reduce market liquidity. The largest reduction in market liquidity stems from an introduction of a short selling ban, while the introduction of a Tobin Tax ranks second. According to the GLS coefficients, a transaction tax of 0.3\%\footnote{See Table~\ref{tab_calibration} on page~\pageref{tab_calibration} in the appendix for the calibration used. Within the regression model all exogenous variables are dummy variables.} reduces the average amount of traded assets of a single  agent by 0.13 per timestep. 
 However, also surprising is the fact that the combined introduction of a transaction tax and a VaR~limit significantly offsets part of the individual liquidity reducing effects.\\

\begin{table}[hb]
\centering
\begin{tabular}{rrrlrrlrrlrrl}

           & {\bf exogenous} & \multicolumn{ 2}{c}{{\bf liquidity}} &     {\bf } & \multicolumn{ 2}{c}{{\bf volatility (sd)}} &     {\bf } & \multicolumn{ 2}{c}{{\bf kurtosis}} &     {\bf } & \multicolumn{ 2}{c}{{\bf defaults }} \\
\hline
           &  Intercept &      0.824 &        *** &            &      2.677 &        *** &            &       2.16 &        *** &            &    310.935 &        *** \\

           &        VaR &     -0.006 &        *** &            &      -0.01 &        *** &            &     -0.186 &        *** &            &     -9.745 &        *** \\

           &      ssban &     -0.133 &        *** &            &     -0.046 &        *** &            &      0.344 &        *** &            &    -84.836 &        *** \\

           &         TT &     -0.112 &        *** &            &      0.022 &        *** &            &       -0.4 &        *** &            &     -3.225 &        *** \\

           & (VaR*ssban) &      0.002 &          . &            &          0 &            &            &      -0.08 &            &            &      3.883 &        *** \\

           &   (VaR*TT) &     -0.001 &          . &            &     -0.003 &          . &            &     -0.008 &            &            &      -1.03 &            \\

           & (ssban*TT) &      0.021 &        *** &            &      0.005 &         ** &            &     -0.165 &         ** &            &      5.635 &        *** \\

           & (VaR*TT*ssban) &          0 &            &            &      0.001 &            &            &       0.06 &            &            &      0.161 &            \\
\hline
           & $adj. R^{2}$ &      0.887 &            &            &      0.158 &            &            &      0.034 &            &            &      0.738 &            \\
\hline
\end{tabular}

Significance Codes: 0 *** 0.001 ** 0.01 * 0.05 . 0.1 
\caption{Results of feasible GLS regression.}
\label{tab_results} 

\end{table}

\begin{figure}
			\begin{center}
			\includegraphics[width=1\textwidth,keepaspectratio]{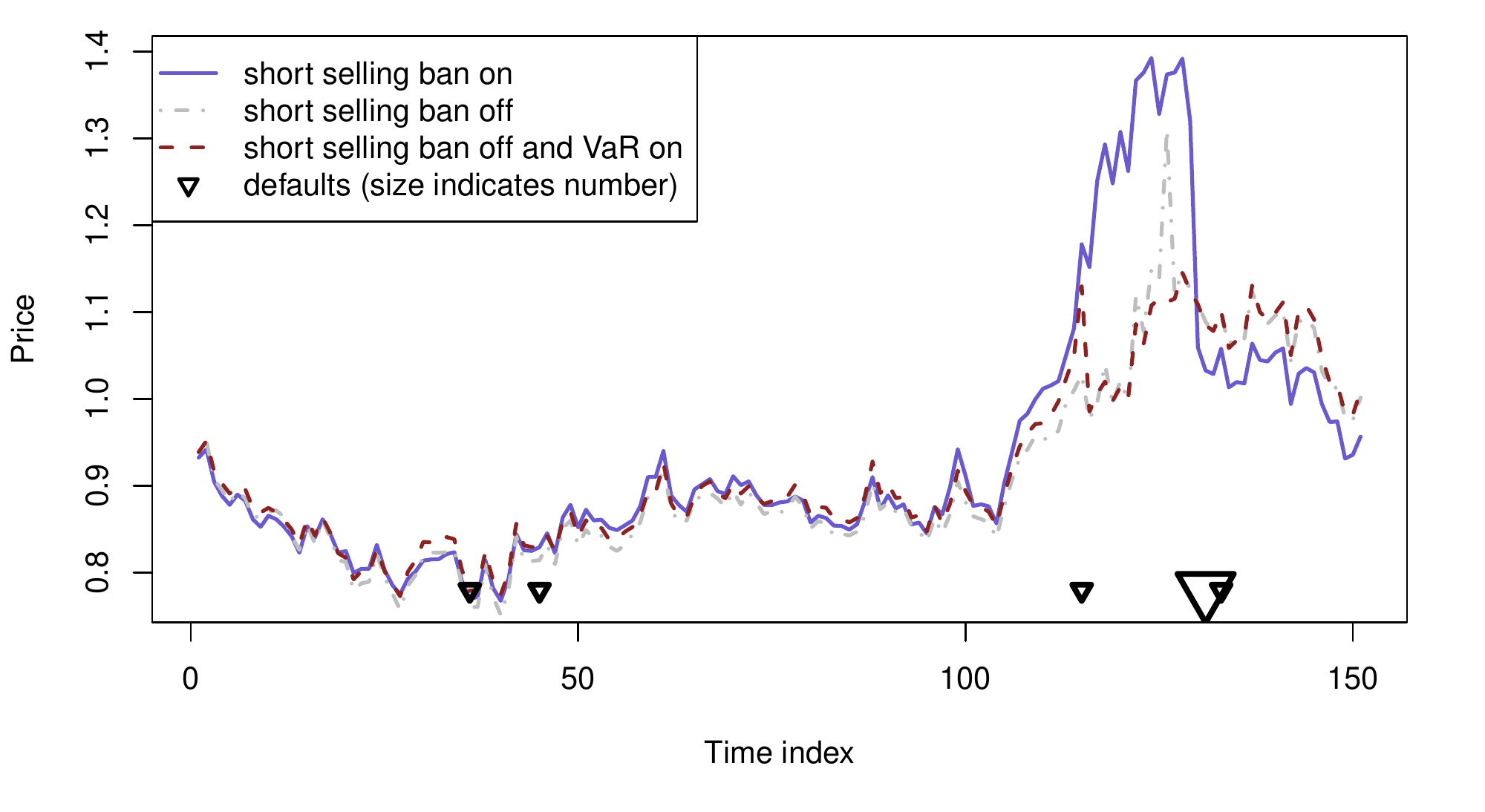}
			\end{center}
				\caption{Price dynamics display higher kurtosis under a short selling ban.}
				\label{pricemovements}
			\end{figure}

 As anticipated from figure~\ref{box1}, both, a short selling ban and a VaR~limit, temper market movements by reducing market volatility. 
 More surprising is probably the fact that a Tobin Tax \textit{increases} market volatility by a statistically significant amount. In line with volatility, obligatory VaR~limits remedies huge swings in markets as seen in the column of kurtosis. 
  Likewise, tail events occur more seldom when a transaction tax is introduced. Both effects are statistically and economically significant. By contrast, a short selling ban positively influences  the probability of market crashes,  via a prior build up of market bubbles. The fact that a short selling ban reduces volatility while increasing the likelihood of tail events  emerges due to the absence of critical investors. Bubbles are nurtured in a calm  environment of low volatility, which lead to crashes when resolved.   Figure~\ref{pricemovements} displays a typical asset price movement under three regimes that visualizes this. At first, there is hardly any difference in the price level and volatility is low. At a certain point, prices start to increase steeply. This is also the point where they start to move away across regimes. While in the VaR regime and even in the unrestricted model price rises are more modest, with short selling prohibited the price rises extraordinarily quickly. 
  The following fall comes certain and costs the default of four agents. Consequent wealth effects cause a lower average price level in the following periods compared to the other regimes. However, regarding  the respective GLS coefficients of the interaction terms in Table~\ref{tab_results}, this dynamic is mitigated, when a short selling ban is combined with VaR~limits or a Tobin Tax.
\\ 

 The number of defaults is negatively associated with all of the regulatory measures. By far the strongest reductions comes from a short selling ban. This is easily explainable: with short selling present in the market, the level of risk is much higher, as agents in a short position would otherwise have no exposure. Additionally, they require a counterpart for their position. Hence, the level of risk across the system is higher, therefore leading to substantially more defaults.\\

\begin{figure}
			\begin{center}
			\includegraphics[width=0.68\textwidth,keepaspectratio]{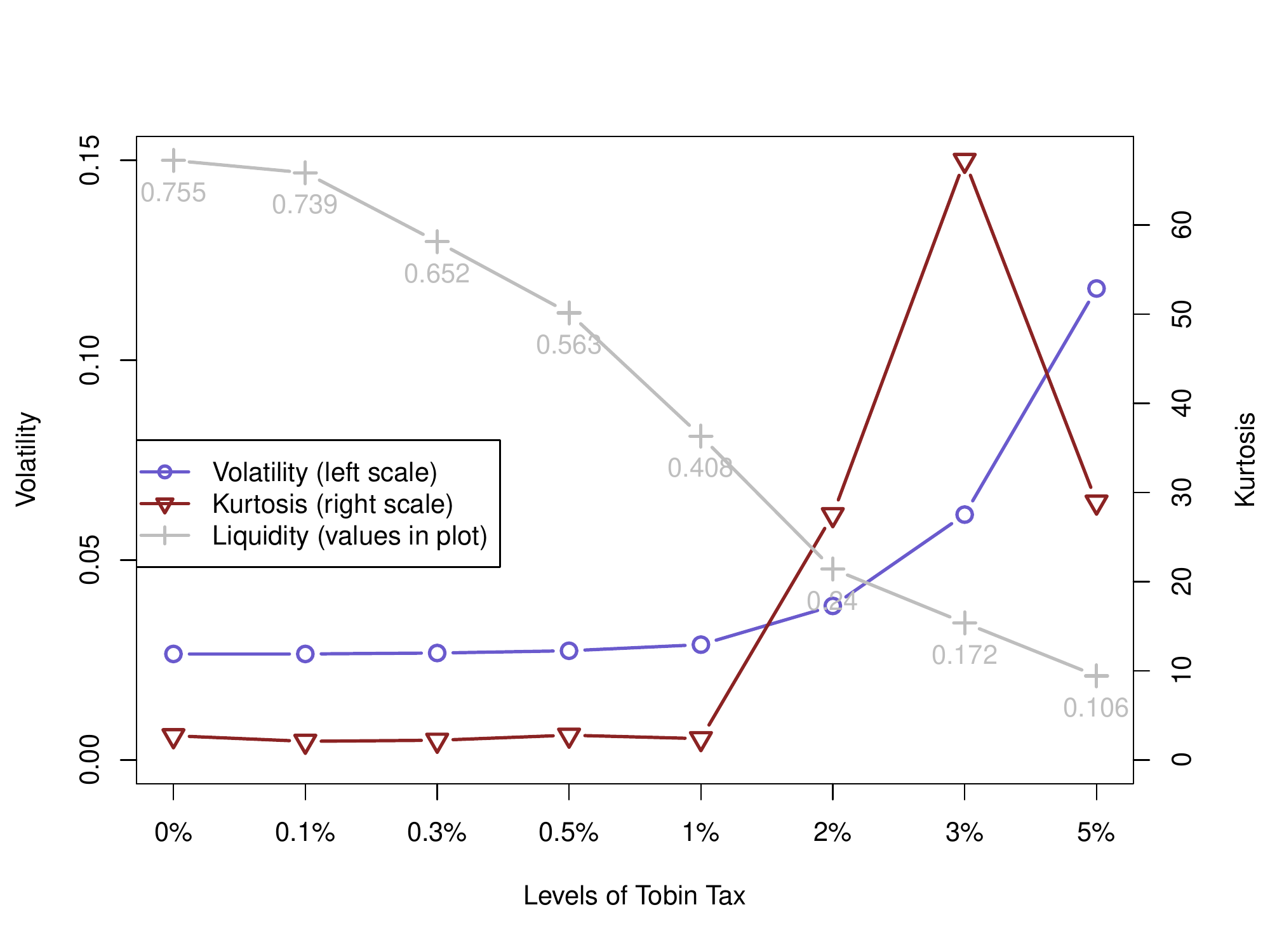}
			\end{center}
\begin{center}
	\footnotesize{The values are means across runs.}		
\end{center}
			\caption{Market characteristics under different levels of Tobin Tax.}
				\label{tobinlevels}
			\end{figure}

While the result that a Tobin Tax increases volatility and reduces tail risk is interesting in itself, one might be interested in how this conclusion changes when the tax level varies (i.e. deviates from its standard value of 0.003). Indeed, e.g. \cite{Westerhoff-2003} finds a dependence of the results on the level of the tax. Consequently, simulations with a Tobin Tax of 0.1\%, up to a level of 5\% were run. Table~\ref{tab_tobin1} and Figure~\ref{tobinlevels} display the results. While there is only a modest increase in volatility noticeable up to a level of 1\%, volatility increases quite drastically above 1\%. At 5\% average market volatility outreaches 10\%, a substaintail increase from its  inital value. As already noted above, a Tobin Tax has a mitigating effect on tail risk in the model applied. This can also be seen  in Table~\ref{tab_tobin1}, where the kurtosis of returns is reduced by a Tobin Tax. However, there is --- indeed --- a certain threshold, at which the medicine is overdosed. At a level of 2\% the average kurtosis jumps and market stability is figuartively dead at any higher level of the tax.\footnote{
For a display of results across different regulatory regimes see  Table~\ref{tab_tobin2} in the appendix, page~\pageref{tab_tobin2}.}\\

\begin{table}[hb]
\centering
\begin{tabular}{lrrrrrrrr}
\hline
{\bf Level of} &            &            &            &            &            &            &            &            \\

{\bf Tobin Tax} &          0 &      0.001 &      0.003 &      0.005 &       0.01 &       0.02 &       0.03 &       0.05 \\
\hline
{\bf Volatility (sd) } &            &            &            &            &            &            &            &            \\

      mean &     0.0265 &     0.0265 &     0.0268 &     0.0273 &     0.0288 &     0.0385 &     0.0614 &     0.1179 \\

    median &     0.0265 &     0.0265 &     0.0267 &     0.0272 &     0.0287 &     0.0310 &     0.0361 &     0.0458 \\

75\%-quantile &     0.0269 &     0.0269 &     0.0271 &     0.0276 &     0.0291 &     0.0315 &     0.0848 &     0.1755 \\
\hline
{\bf Kurtosis } &            &            &            &            &            &            &            &            \\

      mean &     2.7001 &     2.0776 &     2.1974 &     2.7656 &     2.3772 &    27.5035 &    67.2611 &    28.9119 \\

    median &     1.8470 &     1.7456 &     1.3924 &     1.0233 &     0.9122 &     1.9334 &     6.2367 &     6.3635 \\

75\%-quantile &     2.8359 &     2.7072 &     2.2074 &     1.6381 &     1.1505 &     2.2974 &     $>$100 &     $>$100 \\
\hline
\end{tabular}  
\caption{Statistics of returns under different levels of Tobin Tax. }
\label{tab_tobin1} 

\end{table}

What can we now learn from the results? First,  the results show that in the chosen setting a mandatory risk limit is the only measure that is beneficial from all perspectives\footnote{ Liquidity is omitted here.}. A ban on short selling  reduces market volatility, but comes with an increase in tail risk. On the other hand, a Tobin Tax  reduces the occurrence of bubbles while at the same time makes markets more volatile. However, when increased over a certain thresholds results turn and a Tobin Tax clearly contributes to market instability from all perspectives.\\     

Second, the interplay of measures does play a role in judging on the regulatory medicine to be prescribed. When a mandatory risk limit or a Tobin Tax is present, a ban of short selling has significantly lower impact on tail risk than without. The column concerning the number of defaults in Table~\ref{tab_results} also indicates, that regulatory measures can block each other to some extent. While the interplay of these measures should not be left unconsidered when deciding on their implementation, there is no evidence that they turn individual effects in a different direction.\\

However, when interpreting the results one should bare in mind that the model setup is an abstraction and that while it provides a range of features, certain  shortcomings should be addressed in future research:
The high dimensionality of input parameters brings about the question, if the results are stable to a different calibration. Tests in this direction are already on the way. 
Similarly, one could ask if there would be a change in results when the  risk limits are only applied by the largest agents or if their conception is more homogeneous across agents.  Likewise, the question arises what if not the whole market would be subjected to a Tobin Tax but tax havens are present. Furthermore, with a short selling ban in place market participants might anticipate the absence of  short sellers  and incorporate it in their demand decisions, thus mitigating the risk of bubbles.

\section{Conclusions} \label{conclusions}

This paper introduces an artificial market where  agents trade a single asset. The conception of relatively rational agents allows for a straight forward implementation of regulatory measures. These are a short selling ban, a Tobin Tax, a mandatory Value-at-Risk~limit and any arbitrary combination of these. In its unregulated version, the model is capable of reproducing stylised facts of financial markets, most notably fat tails and clustered volatility.\\ 

Introducing regulatory measures constitutes an intervention into a complex system, whose consequences, side effects and joint interplay are ex ante unclear. 
The results described in Section~\ref{results} constitute a reduction of market liquidity under each of the regulatory regimes. A finding less surprising than the one concerning 
market stability: the results indicate that only a mandatory risk limit is beneficial from every perspective, 
while a short selling ban --- though reducing volatility --- increases tail risk. The contrary holds true for a Tobin Tax: it reduces the occurrence of crashes but increases  volatility --- an outcome that shows the importance of prior testing. However, when increased over a certain thresholds results turn and a Tobin Tax clearly contributes to market instability from all perspectives. Furthermore, the interplay of measures is not negligible. Regression analyses show that measures can block each other and a well chosen combination can mitigate unforeseen side effects. \\
 
However, further research is indeed needed to test the robustness of the results with regard to the calibration.
The high dimensionality of input parameters makes this a challenging task, but a worthy considering the  necessity for prior testing, as Dirk Helbing would argue.

\begin{appendix}
\section{Appendix} 
\label{appendix}


Table~\ref{tab_calibration} presents the values used for simulation. The model was calibrated to fit roughly weekly data of stock markets. Where possible values from \cite{thurner-2009} were used. Each run composes 4000 timesteps. One draw of $\vec{\epsilon_{t}}$  was 
 used for every regulatory regime in sequence. \\

Table~\ref{tab_quantreg} depicts the analogon to Table~\ref{tab_results}, but instead of feasible GLS quantile regression is used to model the medians of the market characteristics. Standard errors of coefficients were obtained using bootstrapping methods. The values are strikingly close to the ones of the feasible GLS regression.\\

Table~\ref{tab_tobin2} shows the results of a varying degree of Tobin Tax and varying regulatory regimes in place (i.e. a short stelling ban and/or a mandatory risk limit). 
For the aggregated view (across all regimes) see Table~\ref{tab_tobin1} in Section~\ref{results}, page~\pageref{tab_tobin1}. \\

The simulation was set up in R programming language. 

\begin{sidewaystable}[hb]
\centering
\begin{tabular}{lllrr}

{\bf Parameter} & {\bf Comment} &  {\bf See} & {\bf Values} \\
\hline
\hline
   $N^{a}$ & number of agents & Equ.~(\ref{eq-price}) &        150 \\
\hline
   $N^{s}$ & numer of assets & Equ.~(\ref{eq-price}) &  $N^{a}*3$ \\
\hline
$\beta_{i}$ & agressiveness of agents & Equ.~(\ref{eq-demand}) &      10-50 \\
\hline
$W^{}_{i,0}$ & initial wealth of funds & Equ.~(\ref{eq-wealth}) &          2 \\
\hline
$\lambda^{max}$ & maximum leverage & Equ.~(\ref{eq-demand}) &         10 \\
\hline
   $ \tau$ & fraction of agents never & Footn.~(\ref{f-tau}) &       0,95 \\

           & \hspace{0.2cm} taking short positions &            &            \\
\hline
    $\rho$ & persistence of perc.  & Equ.~(\ref{eq-perc}) &       0.99 \\

           & \hspace{0.2cm} fundamental values &            &            \\
\hline
       $V$ & fundamental value & Equ.~(\ref{eq-perc}) &          1 \\
\hline
$\Sigma^{2}$ & covariance matrix for perc. & Equ.~(\ref{eq-perc}) & diag. elements $0.025^{2}$,  \\

           & \hspace{0.2cm}  fundamental values &            & $\quad$ off-diag. $0.4 * (0.025)^{2}$ \\
\hline
$\mu_{i,t}$ & mean of returns for & Equ.~(\ref{eq-var}) & emp. mean of last  \\

           & \hspace{0.2cm}  VaR-calc. &            & $\max(\beta_{i})*10/\beta_{i}$ obs. \\
\hline
$\sigma_{i,t}$ & sd of returns for  & Equ.~(\ref{eq-var}) & emp. sd of last \\

           & \hspace{0.2cm}  VaR-calc. &            & $\max(\beta_{i})*10/\beta_{i}$ obs. \\
\hline
 Tobin Tax &  Tobin Tax & Equ.~(\ref{eq-demand-tobin}) &      0.003 \\
\hline
  $\Gamma$ & Threshold for Tobin Tax & Equ.~(\ref{eq-demand-tobin}) & $\beta_{i}/0.14*\,$Tobin Tax \\
\hline
\end{tabular}

\caption{Values used for simulation.}
\label{tab_calibration} 

\end{sidewaystable}

\begin{table}[hb]
\centering
\begin{tabular}{rrrlrrlrrlrrl}

           & {\bf exogenous} & \multicolumn{ 2}{c}{{\bf liquidity}} &     {\bf } & \multicolumn{ 2}{c}{{\bf volatility (sd)}} &     {\bf } & \multicolumn{ 2}{c}{{\bf kurtosis}} &     {\bf } & \multicolumn{ 2}{c}{{\bf defaults }} \\
\hline
           &  Intercept &      0.821 &        *** &            &      2.676 &        *** &            &      1.821 &        *** &            &        308 &        *** \\

           &        VaR &     -0.006 &        *** &            &      -0.01 &        *** &            &     -0.199 &        *** &            &         -9 &        *** \\

           &      ssban &     -0.131 &        *** &            &     -0.046 &        *** &            &      0.283 &        *** &            &        -83 &        *** \\

           &         TT &     -0.112 &        *** &            &      0.021 &        *** &            &     -0.407 &        *** &            &         -3 &        *** \\

           & (VaR*ssban) &      0.003 &         ** &            &          0 &            &            &     -0.051 &            &            &          3 &         ** \\

           &   (VaR*TT) &     -0.001 &            &            &     -0.004 &          * &            &      0.024 &            &            &         -1 &            \\

           & (ssban*TT) &      0.021 &        *** &            &      0.006 &          * &            &     -0.125 &          * &            &          6 &        *** \\

           & (VaR*TT*ssban) &          0 &            &            &      0.002 &            &            &      0.024 &            &            &          0 &            \\
\hline
\end{tabular}  
Significance Codes: 0 *** 0.001 ** 0.01 * 0.05 . 0.1 
\caption{Results of quantile regression for the median.}
\label{tab_quantreg} 

\end{table}

\begin{table}[hb]
\centering

\begin{tabular}{lrrrrrrrr}
\hline
{\bf Level of} &            &            &            &            &            &            &            &            \\

{\bf Tobin Tax} &          0 &      0.001 &      0.003 &      0.005 &       0.01 &       0.02 &       0.03 &       0.05 \\
\hline
{\bf Volatility (sd)} &            &            &            &            &            &            &            &            \\

           &            &            &         \multicolumn{ 4}{c}{ssban=off \& VaR=off} &            &            \\

      mean &     0.0268 &     0.0268 &     0.0270 &     0.0276 &     0.0291 &     0.0311 &     0.0335 &     0.0424 \\

    median &     0.0268 &     0.0268 &     0.0270 &     0.0275 &     0.0291 &     0.0311 &     0.0335 &     0.0424 \\

75\%-quantile &     0.0271 &     0.0271 &     0.0273 &     0.0279 &     0.0294 &     0.0314 &     0.0340 &     0.0431 \\

           &            &            &          \multicolumn{ 4}{c}{ssban=on \& VaR=off} &            &            \\

      mean &     0.0263 &     0.0263 &     0.0266 &     0.0272 &     0.0288 &     0.0308 &     0.0333 &     0.0424 \\

    median &     0.0263 &     0.0263 &     0.0266 &     0.0272 &     0.0288 &     0.0308 &     0.0333 &     0.0424 \\

75\%-quantile &     0.0267 &     0.0266 &     0.0269 &     0.0275 &     0.0291 &     0.0312 &     0.0338 &     0.0431 \\

           &            &            &          \multicolumn{ 4}{c}{ssban=off \& VaR=on} &            &            \\

      mean &     0.0267 &     0.0267 &     0.0269 &     0.0273 &     0.0288 &     0.0463 &     0.0890 &     0.1915 \\

    median &     0.0267 &     0.0266 &     0.0268 &     0.0273 &     0.0286 &     0.0312 &     0.0836 &     0.1756 \\

75\%-quantile &     0.0270 &     0.0270 &     0.0272 &     0.0277 &     0.0289 &     0.0683 &     0.1025 &     0.2133 \\

           &            &            &           \multicolumn{ 4}{c}{ssban=on \& VaR=on} &            &            \\

      mean &     0.0263 &     0.0262 &     0.0265 &     0.0270 &     0.0285 &     0.0462 &     0.0898 &     0.1967 \\

    median &     0.0262 &     0.0262 &     0.0264 &     0.0270 &     0.0284 &     0.0309 &     0.0859 &     0.1757 \\

75\%-quantile &     0.0265 &     0.0265 &     0.0268 &     0.0273 &     0.0286 &     0.0681 &     0.1015 &     0.2182 \\
\hline
{\bf Kurtosis} &            &            &            &            &            &            &            &            \\

           &            &            &         \multicolumn{ 4}{c}{ssban=off \& VaR=off} &            &            \\

      mean &     2.9987 &     2.0928 &     2.3827 &     4.7605 &     2.4309 &     1.9394 &     3.2479 &     4.9251 \\

    median &     1.8204 &     1.7082 &     1.4136 &     1.0377 &     0.9532 &     1.8795 &     3.2058 &     4.8683 \\

75\%-quantile &     2.7932 &     2.6700 &     2.2109 &     1.6856 &     1.2171 &     2.1333 &     3.5810 &     5.4066 \\

           &            &            &          \multicolumn{ 4}{c}{ssban=on \& VaR=off} &            &            \\

      mean &     2.8150 &     2.2957 &     2.2823 &     3.2105 &     2.7176 &     1.9308 &     3.3086 &     4.9295 \\

    median &     2.1041 &     1.9867 &     1.5727 &     1.1959 &     0.9412 &     1.8772 &     3.2650 &     4.8795 \\

75\%-quantile &     3.1662 &     3.0241 &     2.4672 &     1.8577 &     1.1714 &     2.1391 &     3.6347 &     5.3922 \\

           &            &            &          \multicolumn{ 4}{c}{ssban=off \& VaR=on} &            &            \\

      mean &     2.4366 &     1.8568 &     2.0085 &     1.1941 &     2.2026 &     $>$100 &     $>$100 &     $>$100 \\

    median &     1.6215 &     1.5430 &     1.2379 &     0.8922 &     0.8696 &     1.9902 &     $>$100 &     $>$100 \\

75\%-quantile &     2.5234 &     2.4760 &     1.9687 &     1.4305 &     1.0994 &     $>$100 &     $>$100 &     $>$100 \\

           &            &            &           \multicolumn{ 4}{c}{ssban=on \& VaR=on} &            &            \\

      mean &     2.5490 &     2.0652 &     2.1160 &     1.8971 &     2.1578 &     $>$100 &     $>$100 &     $>$100 \\

    median &     1.8530 &     1.7365 &     1.3694 &     0.9849 &     0.8784 &     1.9899 &     $>$100 &     $>$100 \\

75\%-quantile &     2.8225 &     2.6501 &     2.1724 &     1.4909 &     1.1164 &     $>$100 &     $>$100 &     $>$100 \\
\hline
\end{tabular}  
 
\caption{Statistics of returns under different levels of Tobin Tax.}
\label{tab_tobin2} 

\end{table}

\end{appendix}

 \label{lit}
\bibliography{literatur}


\bibliographystyle{apalike}
\end{document}